\documentclass[showpacs,prl,amsmath,amssymb,superscriptaddress,nobibnotes,twocolumn, preprintnumbers]{revtex4-1}
\usepackage{graphicx}
\usepackage{endnotes}
\usepackage{bm}
\usepackage{epsfig}
\usepackage{amsmath}

\newcommand{\ie}{{\it i.e.}}
\newcommand{\eg}{{\it e.g.}}
\newcommand{\etal}{{\it et al.}}
\newcommand{\SrIr}{Sr$_{3}$Ir$_4$Sn$_{13}$}
\newcommand{\SrRh}{Sr$_{3}$Rh$_4$Sn$_{13}$}

\newcommand{\CaSrRhx}{(Ca$_{x}$Sr$_{1-x}$)$_3$Rh$_4$Sn$_{13}$}
\newcommand{\CaSrRhqcp}{(Ca$_{0.9}$Sr$_{0.1}$)$_3$Rh$_4$Sn$_{13}$}
\newcommand{\CaSrIrx}{(Ca$_{x}$Sr$_{1-x}$)$_3$Ir$_4$Sn$_{13}$}
\newcommand{\CaRh}{Ca$_{3}$Rh$_4$Sn$_{13}$}
\newcommand{\CaIr}{Ca$_{3}$Ir$_4$Sn$_{13}$}

\newcommand{\CaCo}{Ca$_{3}$Co$_4$Sn$_{13}$}
\newcommand{\CaIrCo}{Ca$_3$(Ir$_{0.91}$Co$_{0.09}$)$_4$Sn$_{13}$}
\newcommand{\Tstar}{$T^*$}

\begin{document}

%%%%%%%%%%%%%%%%%%%%% TITLE %%%%%%%%%%%%%%%%%%%
\title{Soft phonon modes in the vicinity of the structural quantum critical point}
%%%%%%%%%%%%%%%%%%%% AUTHORS %%%%%%%%%%%%%%%%%%
\author{Y.~J.~Hu}
\author{Y.~W.~Cheung}
\author{W.~C.~Yu}
\affiliation{Department of Physics, The Chinese University of Hong Kong, Shatin, New Territories, Hong Kong, China}

\author{Masaki~Imai}
\author{Hibiki~Kanagawa}
\author{Joichi~Murakawa}
\affiliation{Department of Chemistry, Graduate School of Science, Kyoto University, Kyoto 606-8502, Japan}
\author{Kazuyoshi~Yoshimura}
\affiliation{Department of Chemistry, Graduate School of Science, Kyoto University, Kyoto 606-8502, Japan}
\affiliation{Research Center for Low Temperature and Materials Sciences, Kyoto University, Kyoto 606-8501, Japan}

\author{Swee~K.~Goh}
\email{skgoh@phy.cuhk.edu.hk}
\affiliation{Department of Physics, The Chinese University of Hong Kong, Shatin, New Territories, Hong Kong, China}
\date{\today}
%\preprint{ver. 2.0}
%%%%%%%%%%%%%%%%%%% ABSTRACT %%%%%%%%%%%%%%%%%%%%

\begin{abstract}

The quasi-skutterudite superconductors $A_3T_4$Sn$_{13}$  ($A$=Sr, Ca; $T$=Ir, Rh, Co) are highly tunable featuring a structural quantum critical point. We construct a temperature-lattice constant phase diagram for these isovalent compounds, establishing \CaRh\ and \CaCo\ as members close to and far away from the structural quantum critical point, respectively. Deconvolution of the lattice specific heat and the electrical resistivity provide an approximate phonon density of states $F(\omega)$ and the electron-phonon transport coupling function $\alpha_{tr}^2F(\omega)$ for \CaRh\ and \CaCo, enabling us to investigate the influence of the structural quantum critical point. Our results support the scenario of phonon softening close to the structural quantum critical point, and explain the enhancement of the coupling strength on approaching structural instability.

\end{abstract}

%\pacs{74.25.fc, 74.25.Bt, 63.20.kd}

%74.25.-q - properties of superconductors
%62.50.-p - high pressure effects in solids and liquids
%74.25.Op - superconductors - critical fields.
%74.40.Kb	Quantum critical phenomena
%74.62.-c	Transition temperature variations, phase diagrams
%65.40.Ba	Heat capacity
%64.70.Tg	Quantum phase transitions

\maketitle

%%%%%%%%%%%%%%%%%%%%% INTRO %%%%%%%%%%%%%%%%%%%%
\section{Introduction}

The investigation of the interplay between structural instability and superconductivity has a long history. Early example includes A-15 compounds Nb$_3$Sn and V$_3$Si in which an enhancement of the superconducting critical temperature ($T_c$) was reported with the suppression of structural transition temperature \cite{Chu1973,Chu1974}. More recently, the material base has been expanded to include transition metal dichalcogenides derived from IrTe$_2$ \cite{Yang2012,Pyon2012,Fang2013,Kamitani2016,Kudo2016} as well as Ni- and Fe-based superconductors \cite{Cruz2008,Yoshizawa2012,Niedziela2011,Kudo2012,Hirai2012}, in which superconductivity emerges at the first-order structural transition boundary. These studies explicitly highlight the role of structural instability on the stabilization of the superconductivitiy.

The stannide superconductors with a chemical composition $A_3T_4$Sn$_{13}$ ($A$=Ca, Sr, La; $T$=Ir, Rh, Co) have recently been studied with a wide range of probes \cite{Klintberg2012,Goh2015,Yu2015,Hou2016, Kuo2014,Kuo2015,Yang2010,Gerber2013,Liu2013,Slebarski2014,Fang2014,BChen2015,Mazzone2015,Wang2012,Biswas2014,Wang2015, Kase2011, Hayamizu2011, Zhou2012, Sarkar2015,Biswas2015, Slebarski2013, Thomas2006, Slebarski2015,Neha2016,Tompsett2014, XChen2015, Cheung2016, Lue2016, Luo2016, Wang2017, Cheung2017}. The superconducting gap symmetry has been established to be of a conventional $s$-wave type \cite{Kase2011,Hayamizu2011,Zhou2012,Wang2012,Biswas2014,Wang2015,Sarkar2015}. In certain compositions, a structural phase transition occurs upon cooling. For instance, \SrIr\ and \SrRh\ with a space group of $Pm\bar{3}n$ at room temperature ($I$ phase) \cite{Kase2011,Klintberg2012} undergo a structural phase transition at $T^*$= 147~K and 138~K respectively, below which superlattice reflections were observed at the $\mathbf{M}$ point, which corresponds to ${\bm q}=(0.5, 0.5, 0)$ and the symmetry equivalents ($I'$ phase) \cite{Goh2015,Lue2016}. The structural transition has been shown to be of second order by the shape of the specific heat jump \cite{Yu2015, Cheung2017}, the absence of hysteresis in resistivity \cite{Goh2015,Klintberg2012, Cheung2017} and a continuous growth of superlattice reflection \cite{Lue2016, Cheung2017} around $T^*$. Crucially, the structural transition temperature \Tstar\ is highly controllable: \Tstar\ can be suppressed to 0~K at a structural quantum critical point (QCP) via a suitable combination of hydrostatic pressure and chemical substitution \cite{Klintberg2012, Goh2015}, giving rise to phase diagrams in which the role of the structural quantum criticality and its influence on the superconductivity can be explored in a systematic manner. 

\CaSrIrx\ \cite{Klintberg2012}, \CaSrRhx\ \cite{Goh2015,Yu2015}, and Ca$_3$(Ir$_{1-y}$Co$_y$)$_4$Sn$_{13}$ \cite{Hou2016} are several substitution series that have been investigated recently. In \CaSrRhx, it has been shown that $T^*$ can be driven to 0~K solely by calcium substitution. In the vicinity of the structural QCP, \ie\ $x_{crit} \approx 0.9$ in \CaSrRhx \cite{Goh2015,Yu2015}, the resistivity is linear in temperature, the Debye temperature is a minimum, $T_c$ takes the maximum value, and the superconducting state is of a strong-coupling nature, as benchmarked by a substantially enhanced gap-to-$T_c$ ratio $2\Delta(0)/k_BT_c$ and normalized specific heat jump $\Delta C/\gamma T_c$ \cite{Carbotte1990,Yu2015}. These observations can all be explained by considering the softening of the relevant phonon mode due to the second-order structural transition. Indeed, calculations have found the softening of phonon modes at the $\mathbf{M}$ point \cite{Tompsett2014, Goh2015}, which was subsequently confirmed by inelastic neutron scattering in \CaIr\ \cite{Mazzone2015}. For \SrIr, phonon softening was observed on approaching $T^*$ from below by ultrafast spectroscopy \cite{Luo2016}.

Recently, Hou \etal\ investigated Ca$_3$(Ir$_{1-y}$Co$_y$)$_4$Sn$_{13}$ \cite{Hou2016} for low Co concentrations ($y\leq0.12$). For this series, \CaIrCo\ appears to be at the part of the phase diagram where \Tstar\ extrapolates to 0~K. Following a well-established method of analyzing specific heat and electrical resistivity \cite{Hou2016,Lortz2006,Lortz2008,Lortz2005,Junod1983,Teyssier2008}, the approximate phonon density of states $F(\omega)$ and the electron-phonon transport coupling function $\alpha_{tr}^2F(\omega)$ of \CaIrCo\ were obtained, leading to the conclusion of phonon-mediated strong-coupling superconductivity, consistent with the observation of Yu \etal\ and Biswas \etal\ in relevant series \cite{Yu2015,Biswas2015}. To gain further insights into the role of structural instability, it is desirable to extend the work of Hou \etal\ to study a composition which is far away from the structural QCP, and this composition should not undergo a structural transition. As we will establish in this manuscript, \CaCo\ ($T_c=6.0$~K from resistivity, space group $Pm\bar{3}n$) is a good candidate. In this work, we report $F(\omega)$ and $\alpha_{tr}^2F(\omega)$ of \CaCo, and for a comparative study, of \CaRh\ which is very close to the structural QCP. Both \CaCo\ and \CaRh\ are in the $I$ phase with no structural transition observed down to the lowest attainable temperature.

%%%%%%%%%%%%%%%%% EXPERIMENTAL %%%%%%%%%%%%%%%%%%%%
\section{Method}

Single crystals of \CaRh\ and \CaCo\ were synthesized by a tin flux method following similar steps as described elsewhere \cite{Yang2010}. The heat capacity was measured using a standard pulse relaxation method. The mass of the \CaRh\ and \CaCo\ are 24.72~mg and 1.08~mg, respectively. Electrical resistivity was measured using the four-contact method. The low temperature and high magnetic field environment were provided by a Physical Property Measurement System (Quantum Design). 

%%%%%%%%%%%%%%%%%%%%% RESULTS %%%%%%%%%%%%%%%%%%
\section{Results and Discussion}

The temperature-pressure phase diagrams constructed for \CaSrRhx\ \cite{Goh2015} and \CaSrIrx\ \cite{Klintberg2012} have established the role of Ca as a provider of chemical pressure in both isovalent substitution series. Furthermore, the two phase diagrams bear a close resemblance, hinting at more universal tuning parameters. Inspired by these observations, we plot in Fig.~1 the $T_c$ and \Tstar\ of the two series against their room-temperature lattice constants. Since Co, Rh and Ir are from the same group in the periodic table, the chemical substitution of the $T$ site in $A_3T_4$Sn$_{13}$ with these elements is also isovalent. Therefore, it is natural to include Ca$_3$(Ir$_{0.91}$Co$_{0.09}$)$_4$Sn$_{13}$ and \CaCo\ in the phase diagram. From Fig.~1, it is immediately clear that \CaCo\ is far away from the structural QCP. Moreover, $T_c$ decreases under pressure with an initial slope $dT_c/dp\sim-$0.4~K/GPa \cite{Logg2014}, which follows the trend of $T_c$ on this part of the phase diagram. Therefore, \CaCo\ is an ideal composition to investigate the right hand part of the phase diagram. Aliovalent stannides such as La$_3$Co$_4$Sn$_{13}$ are excluded from this analysis.
%%%%%%%%%%%%%%%%Figure 1
\begin{figure}[!t]\centering
      \resizebox{9cm}{!}{
              \includegraphics{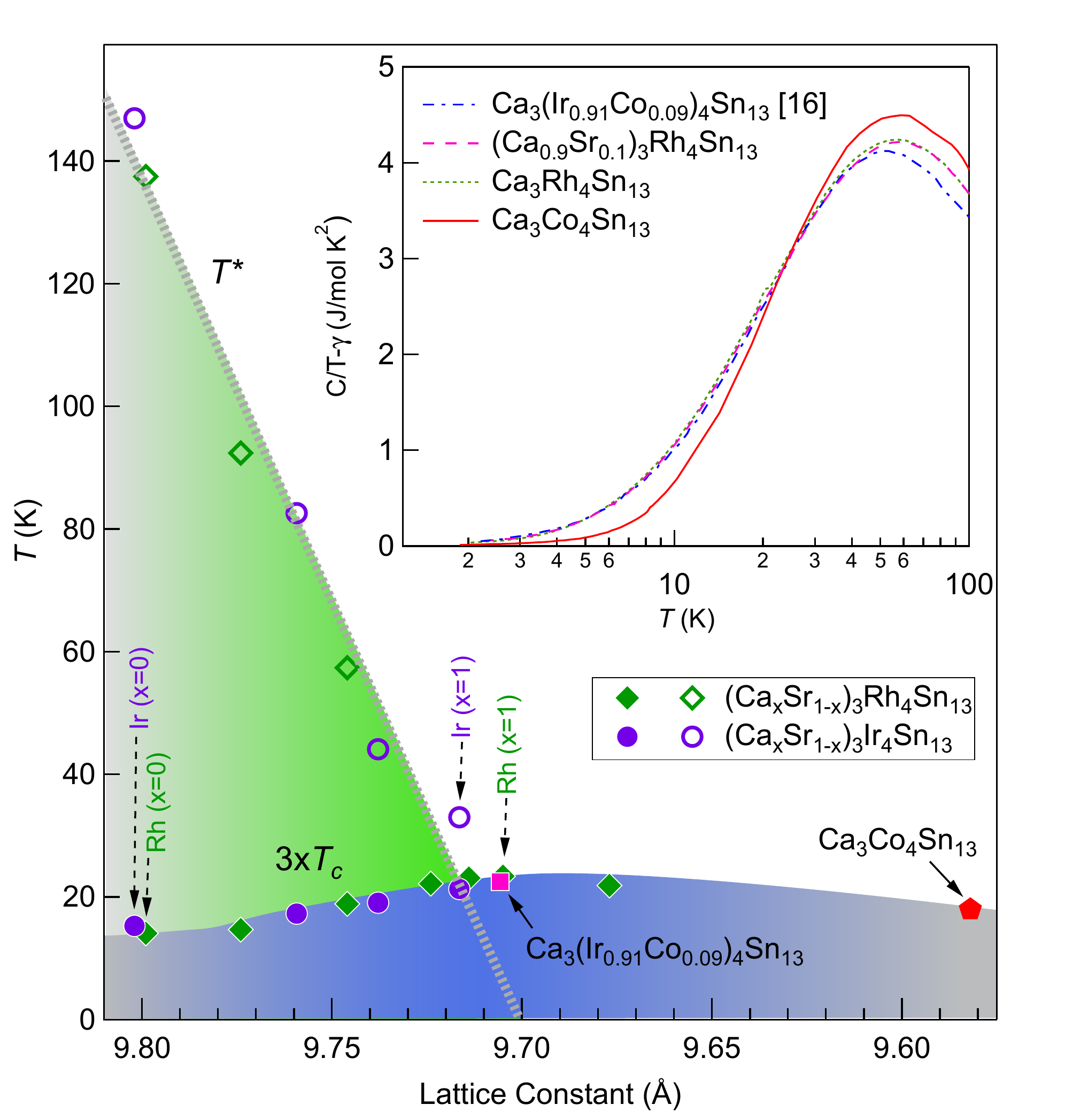}} 	
              \caption{\label{fig1} (Color online) Universal phase diagram with $T_c$ (solid symbols) and \Tstar\ (open symbols) against the room temperature lattice constant. The \CaSrRhx\ series contains $x = 0, 0.25, 0.5, 0.75, 0.9, {\rm and}~1$ and \CaRh\ at 20.6~kbar, and the \CaSrIrx\ series contains $x = 0, 0.5, 0.75, {\rm and}~1$, where a larger $x$ of a given series corresponds to a smaller lattice constant. The end compounds are indicated by arrows. The solid square and solid pentagon denote $T_c$ of \CaIrCo\ and \CaCo, respectively. All transition temperatures are determined from resistivity measurements \cite{Goh2015, Klintberg2012, Hou2016}. The lattice constants of \CaRh\ at 20.6~kbar and the intermediate compounds \CaSrIrx\ and \CaIrCo\ are extrapolated using the lattice constant of respective end compounds according to Vegard's law. The lattice constants of \CaSrRhx\ are available from Ref.~\cite{Goh2015}. The dashed `\Tstar-line' is a guide to the eyes. Inset: $(C/T-\gamma)$ against $T$ for \CaCo, \CaRh, \CaSrRhqcp, and \CaIrCo. The data for \CaIrCo\ were digitized from Ref.~\cite{Hou2016}.}
\end{figure}

%%%%%%%%%%%%%%%%Figure 2
\begin{figure}[!t]\centering
       \resizebox{8.5cm}{!}{
              \includegraphics{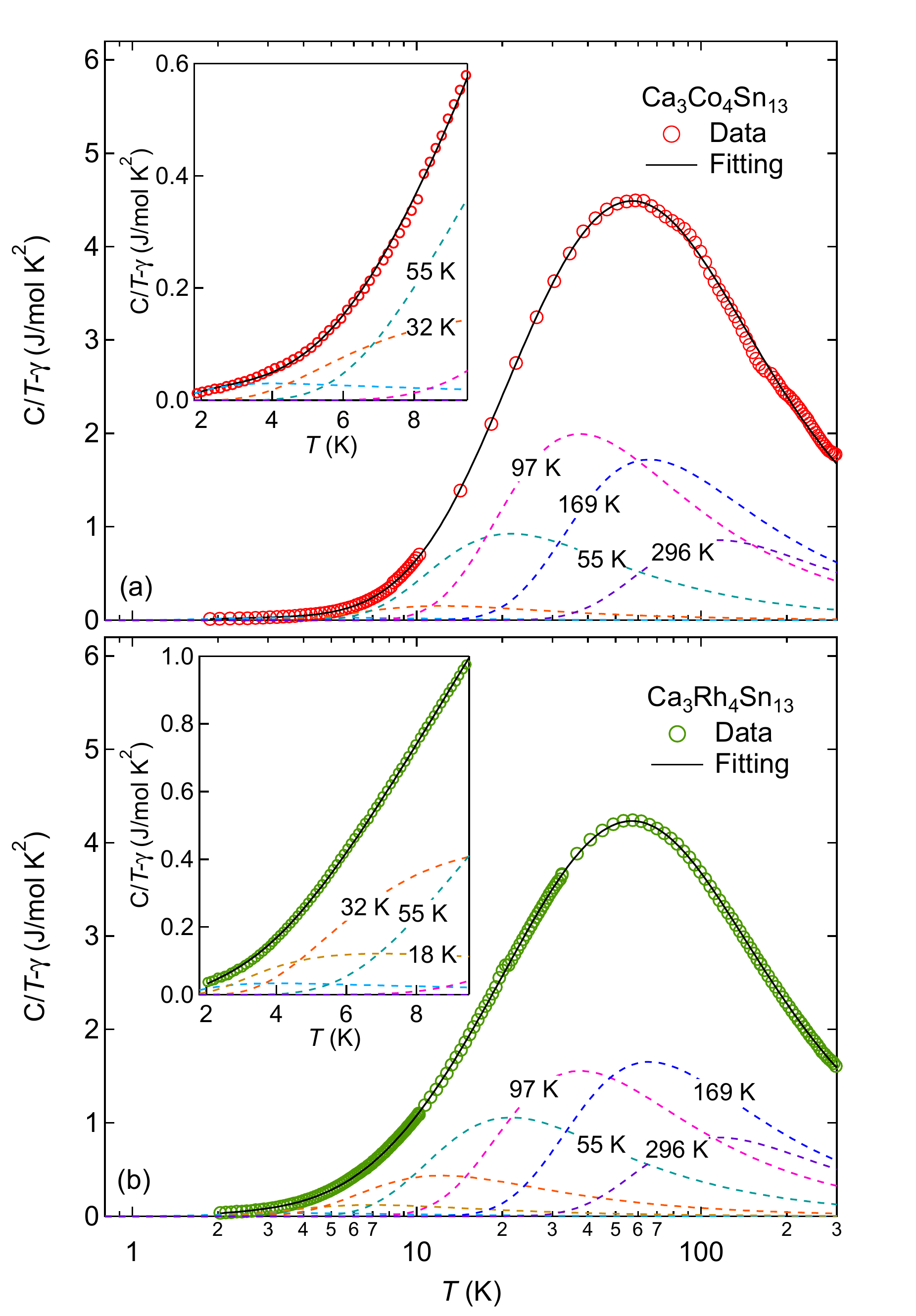}}                				
              \caption{\label{fig2} (Color online) The temperature dependence of the lattice specific heat divided by temperature for (a) \CaCo\ and (b) \CaRh, with insets showing the low temperature part. The solid curve is the fit using Eq.~(1) and the dashed curves are the constituent Einstein components, labelled by the appropriate Einstein temperatures $\Theta_i$. For reference, $\gamma$ is 60.9~mJK$^{-2}$mol$^{-1}$ and 57.2~mJK$^{-2}$mol$^{-1}$ for \CaCo\ and \CaRh, respectively. %For reference, $(\gamma,\Theta_D)$ is (60.9~mJK$^{-2}$mol$^{-1}$, 225~K) and (57.2~mJK$^{-2}$mol$^{-1}$, 164~K) for \CaCo\ and \CaRh, respectively, where $\Theta_D$ is the Debye temperature. %By fitting the peak of the data the 169~K mode is first determined and then expand to a series by $\Theta_{i+1}=1.75\Theta_i$.  
              }
\end{figure}
%%%%%%%%%%%%%%%%%%%%

The normal state specific heat of \CaSrRhqcp, \CaRh, and \CaCo\ were measured from 2~K to 300~K. The Sommerfeld coefficient $\gamma$ was first extracted from the specific heat at low temperature following the standard procedure (\eg\ as described in detail in Ref.~\cite{Yu2015} for both \CaSrRhqcp\ and \CaRh).  This allows us to subtract the electronic contribution from the total heat capacity. In the inset of Fig.~1, the phonon contribution to the specific heat divided by temperature, $(C/T-\gamma)$, is plotted. Additionally, the data for \CaIrCo\ were digitized from Ref.~\cite{Hou2016} for comparison. It is clear that \CaCo\ behaves differently from the others at low temperature. Note that we have avoided the compositions with structural transition, so that no Fermi surface reconstruction occurs and $\gamma$ can be regarded as temperature independent.

For quantitative comparison, we represent the phonon density of states $F(\omega)$ using a basis of Einstein modes: $F(\omega)=\sum_i F_i\delta(\omega-\omega_i)$ with $\hbar\omega_i=k_B\Theta_i$, where $\Theta_i$ and $F_i$ are the Einstein temperature and the weight of the $i^{th}$ Einstein component, respectively. The corresponding specific heat is \cite{Hou2016,Lortz2006,Lortz2008,Lortz2005,Junod1983,Teyssier2008}
\begin{equation}
C_{ph}(T)=N_{A}k_B\sum_i F_i\frac{x_i^2e^{x_i}}{(e^{x_i}-1)^2}
\label{eq:Cph}
\end{equation}
where $x_i=\Theta_i/T$, $N_A$ is the Avogadro's number, and $k_B$ is the Boltzmann constant. Fig.~2 presents the results for \CaCo\ and \CaRh, and the insets display a close-up at low temperature. With seven Einstein modes, equally spaced in the logarithmic $\omega$ scale such that $\omega_{i+1}/\omega_i=\Theta_{i+1}/\Theta_i=$1.75, where the first term $\Theta_1=10.3$~K, we successfully describe the lattice specific heat of both \CaCo\ and \CaRh\ over the entire temperature range, with a sufficient resolution to identify the key difference between the two compounds.

%%%%%%%%%%%%%%%%Figure 3
\begin{figure}[!t]\centering
       \resizebox{8cm}{!}{
              \includegraphics%[trim=3cm 0cm 3cm 1.2cm, clip=true]
              {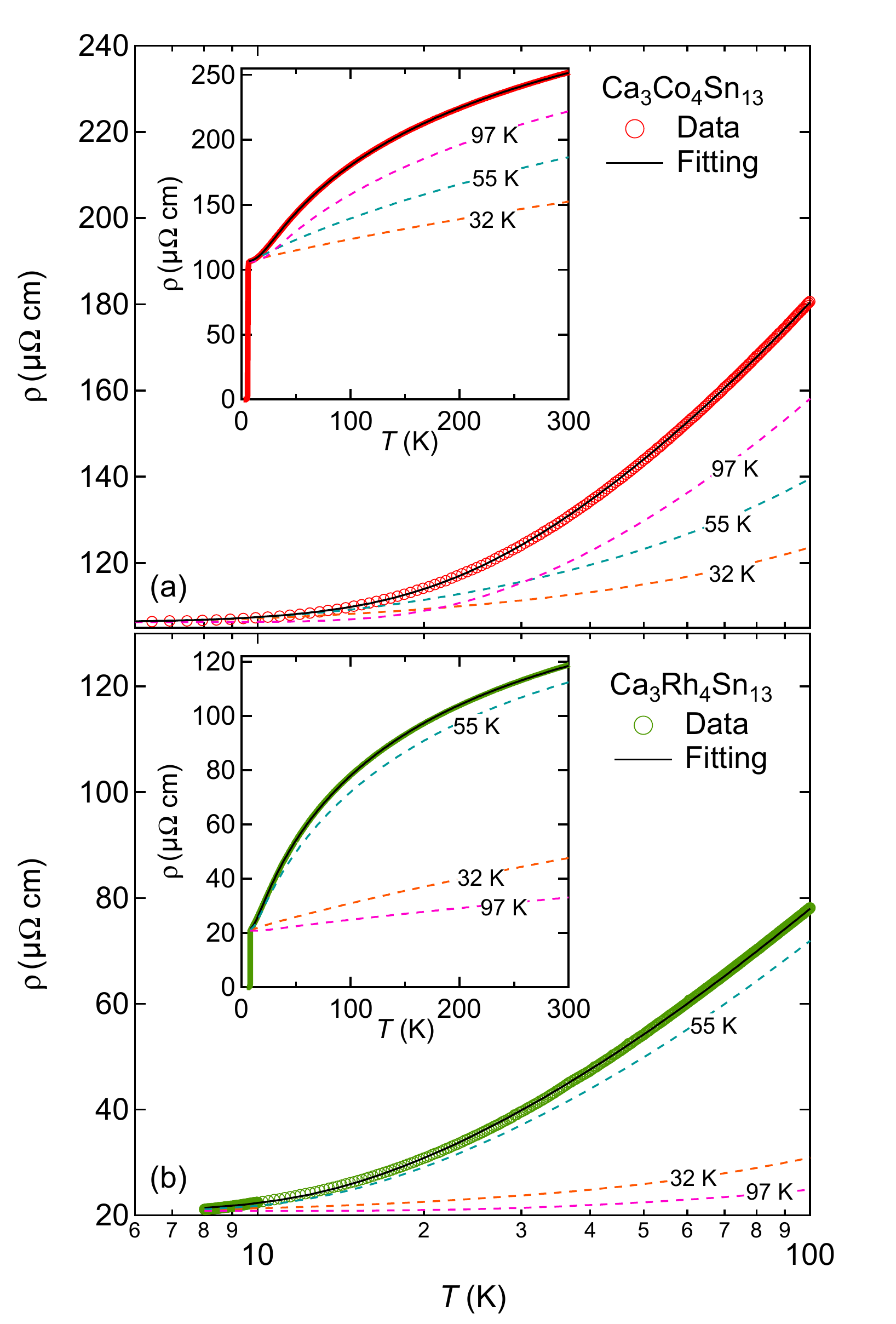}}              				
              \caption{\label{fig3} (Color online) Resistivity data of the entire temperature range and the decomposition of the normal state part for (a) \CaCo\ and (b) \CaRh. The Einstein modes $\Theta_{i+1}=1.75\Theta_i$ are the same as the ones used for the specific heat analysis, and their values are indicated next to the corresponding dashed curves.}
\end{figure}

In an analogous manner, we can study the spectral electron-phonon transport coupling function $\alpha_{tr}^2F(\omega)=\frac{1}{2}\sum_i\lambda_i\omega_i\delta(\omega-\omega_i)$, where $\lambda_i$ is dimensionless, from the decomposition of the total electrical resistivity into discrete components which can be described by the Bloch-Gr\"{u}neisen formula \cite{Hou2016,Lortz2006,Lortz2008,Lortz2005,Junod1983,Teyssier2008}: 
\begin{equation}
\rho_{BG}(T)=\rho(0)+\frac{2\pi}{\epsilon_0\Omega_p^2}\sum_i \lambda_i\omega_i\frac{x_ie^{x_i}}{(e^{x_i}-1)^2}
\label{eq:rhoBG}
\end{equation}
where $x_i=\Theta_i/T$, $\Omega_p$ is the plasma frequency, and $\epsilon_0$ is the dielectric constant. The same set of Einstein modes used in specific heat analysis are employed for the analysis of the resistivity. Above $\sim$50~K, the resistivity starts to exhibit a negative curvature, suggesting a saturation behaviour at higher temperature. Following the empirical parallel-resistor model \cite{Wiesmann1977}, which was developed when the system approaches the Mott limit \cite{Adler1978,Mott1977}, our measured resistivity $\rho(T)$ is analyzed with the following equation,
\begin{equation}
\frac{1}{\rho(T)}=\frac{1}{\rho_{BG}(T)}+\frac{1}{\rho_{sat}},
\label{eq:rhoT}
\end{equation}
where $\rho_{sat}$ is the fitted saturation resistivity. The value of $\rho_{sat}$ is 368~$\mu\Omega$cm and 169~$\mu\Omega$cm for \CaCo\ and \CaRh, respectively. The results are shown in Fig.~3. 

For both \CaCo\ and \CaRh, there are primarily three Bloch-Gr\"{u}neisen components contributing to the resistivity. With the plasma frequency unknown, only the relative weight of $\lambda_i$ can be obtained from the fitting. However,  the electron-phonon coupling constant, $\lambda_{ep}$, can be written as $\lambda_{ep}=\sum_i\lambda_i$. From Ref.~\cite{Hayamizu2011}, $\lambda_{ep}=1.17$ and 1.62 for \CaCo\ and \CaRh, respectively. Hence, the absolute value of $\lambda_i$, and consequently $\alpha_{tr}^2F(\omega)$ can be obtained. 

%%%%%%%%%%%%%%%%Figure 4
\begin{figure}[!t]\centering
       \resizebox{8.5cm}{!}{
              \includegraphics{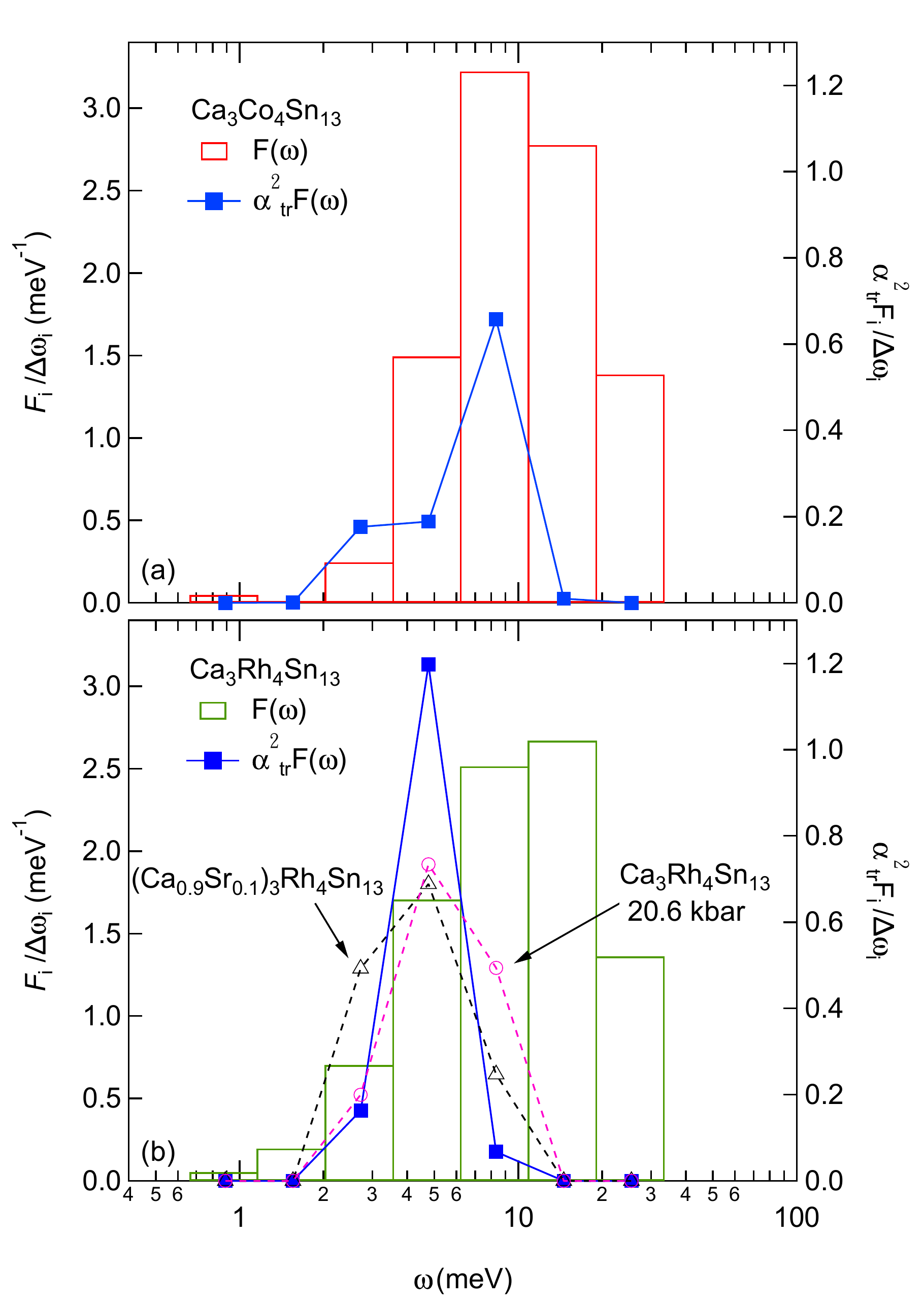} 
                 }           				
              \caption{\label{fig4} (Color online) Phonon density of states per formula unit $F(\omega)$ and the electron-phonon transport coupling function $\alpha_{tr}^2F(\omega)$ for (a) \CaCo\ and (b) \CaRh. The width of the histogram is $\Delta\omega_i=1.75^{0.5}\omega_i-\omega_i/1.75^{0.5}=0.567\omega_i$. Therefore, $F_i/\Delta\omega_i\sim F_i/\omega_i$ and $\alpha_{tr}^2F_i/\Delta\omega_i\sim \lambda_i$. In (b) the open circles and triangles denote $\alpha_{tr}^2F(\omega)$ for \CaRh\ at 20.6~kbar and \CaSrRhqcp, respectively. For clarity, the data for \CaRh\ at 20.6~kbar and \CaSrRhqcp\ have been rescaled.}
\end{figure}

The extracted $\alpha_{tr}^2F(\omega)$ and $F(\omega)$ are plotted in Fig.~4, which provides key insight to the understanding of this material system. Comparing first $F(\omega)$ obtained for both compounds whose weight distribution is represented as a histogram, it is clear that there are more low-energy phonon modes in \CaRh\ than in \CaCo. For example, while the contribution of the $\omega_2=1.55$~meV ($\Theta_2=18.0$~K) mode is negligible in \CaCo, it is finite in \CaRh. Note that $\sum F_i$ for both \CaCo\ and \CaRh\ is close to 60 (Table I), which is the expected total number of phonon modes per formula unit since there are 20 atoms in each formula unit. This verifies the reliability of our data and the accuracy of our analysis. By inspecting ($C/T-\gamma$) for \CaSrRhqcp\ and \CaIrCo\ (inset of Fig.~1), we can expect very similar $F(\omega)$ to that of \CaRh. These results reveal that part of the spectra weight of higher energy modes intrinsically transfer to lower energy as one tunes towards the structural QCP. 

\begin{table}[]
\centering
\label{my-label}
\begin{ruledtabular}
\begin{tabular}{cc|cc|ccc|cc}
\ $\Theta_i$ & &  $\omega_i$ & &  \multicolumn{2}{c}{\CaCo}  && \multicolumn{2}{c}{\CaRh}\\
\ (K) && (meV)  &  & $F_i/\omega_i $ & $\lambda_i$ && $F_i/\omega_i $ & $\lambda_i$ \\
\hline
\ 10.3 && 0.89  & & 0.03  & 0  && 0.03 & 0\\
\ 18.0 && 1.55 & &  0  & 0  && 0.11 & 0\\
\ 31.5 && 2.72 & &  0.14  & 0.20  && 0.40 & 0.19\\
\ 55.2 && 4.76 & &  0.85  & 0.21  && 0.97 & 1.35\\
\ 96.6 && 8.32 & &  1.83  & 0.75  && 1.43 & 0.08\\
\ 169 && 14.6 & &  1.58  & 0.01  && 1.51 & 0\\
\ 296 && 25.5 & &   0.79 & 0  && 0.77 & 0\\
%$\overline{\omega}_{ln}$ & 10.21 & 6.12  && 9.10 & 4.58 &\\
%$<\overline{\omega}^2>^{1/2}$ & 13.81 & 7.12 && 13.47 & 4.81 &\\
\hline
\multicolumn{3}{c}{$\sum \alpha_{tr}^2F_i$} && \multicolumn{2}{c}{3.96} && \multicolumn{2}{c}{3.80}\\
\multicolumn{3}{c}{$\lambda_{ep}$~\cite{Hayamizu2011}}  && \multicolumn{2}{c}{1.17} &&  \multicolumn{2}{c}{1.62}\\
\multicolumn{3}{c}{$\sum F_i$} && \multicolumn{2}{c}{62.63} && \multicolumn{2}{c}{59.54}\\
\end{tabular}
\end{ruledtabular}
\caption{The result of the deconvolution of specific heat and resistivity. The phonon energy series is given by $\omega_{i+1}=1.75\omega_i$. The contribution to $\lambda_{ep}$ from each mode is $\lambda_i=2\alpha_{tr}^2 F_i/\omega_i$, which is determined by using the resultant fitting parameters together with electron-phonon coupling constant $\lambda_{ep}$ from Ref \cite{Hayamizu2011}. The total number of phonon modes per formula unit is given by $\sum F_i$.}
\end{table}

Turning to $\alpha_{tr}^2F(\omega)$, it can be seen that the electrical resistivity of \CaRh\ is dominated by a mode at 4.76~meV (55~K). This is in contrast to the situation in \CaCo\ where modes at higher energies play a more significant role in its electrical transport.  In Table I, we tabulate the numerical values of the key parameters extracted from our analysis. Comparing with \CaCo, \CaRh\ has a smaller $\sum\alpha_{tr}^2F_i$ but larger $\lambda_{ep}$, which is $\sum\alpha_{tr}^2F_i/\omega_i$. This demonstrates how the coupling strength can be enhanced through coupling to soft phonon modes. In fact, linear resistivity has been reported below 50~K for \CaSrRhqcp, which was established to be at the QCP, due to the coupling of the electron and the soft modes \cite{Goh2015}. Similar behaviour was observed in \CaIr\ at 18 kbar \cite{Klintberg2012}. Interestingly, \CaIrCo\ does not show a linear resistivity at low temperature even though it is located near the structural QCP. There, a large contribution from a mode at 12~meV was detected, together with a weaker contribution of a mode at $\sim$~4~meV \cite{Hou2016}. To find out if other modes are softened on approaching structural instability in Ca$_3$(Ir$_{1-y}$Co$_y$)$_4$Sn$_{13}$, it is important to investigate more members of the series, \eg\ $y=0.12$.

To further strengthen the claim that enhancement in coupling strength for \CaRh\ is due to the coupling to the soft mode, we investigate the scaled $\alpha_{tr}^2F(\omega)$ of \CaRh\ at 20.6~kbar and \CaSrRhqcp\ for comparison. As shown in Fig.~4(b), \CaSrRhqcp, which is the closest to the QCP, exhibits an overall softening of $\alpha_{tr}^2F(\omega)$ (open triangles) when compared with the case of \CaRh. Conversely, \CaRh\ at 20.6~kbar, which locates away from the QCP, shows an overall hardening of $\alpha_{tr}^2F(\omega)$. The systematic change in $\alpha_{tr}^2F(\omega)$ highlights the importance of coupling to the phonon modes that are become softer as the system is tuned closer to the structural QCP.

%\section{Summary}
In summary, we have established the universal phase diagram of the isovalent substitution series $A_3T_4$Sn$_{13}$  ($A$=Sr, Ca; $T$=Ir, Rh, Co), and probed the influence of the structural QCP. Through our specific heat and electrical resistivity data, we extracted the phonon density of states and the electron-phonon transport coupling function, and directly compared these parameters in \CaRh\ and \CaCo, where the former compound is close to while the latter is far from the structural QCP. Near the QCP, an enhanced coupling to the low-lying phonon modes is clearly observed. Our work provides key support to models based on soft phonon modes that have been proposed for explaining the novel properties observed in these systems thus far.

\begin{acknowledgments} This work was supported by National Science Foundation China (No. 11504310), Research Grant Council of Hong Kong (ECS/24300214), CUHK Direct Grant (No. 3132719, No. 3132720), CUHK Startup (No. 4930048), and Grant-in-Aids for Scientific Research from Japan Society for the Promotion of Science (No. 16H04131).
\end{acknowledgments}

%\begin{eqnarray}
%	\xi_{ab}^2=\Phi_0/2\pi H_{c2}^c \\
%	\xi_{ab}\xi_c=\Phi_0/2\pi H_{c2}^{ab}
%\end{eqnarray}

%%%%%%%%%%%%%%%%%% BIBLIOGRAPHY USING BIBTEX%%%%%%%%%%%%%%%%%%%%
%\bibliography{refbase}
%\bibliographystyle{h-physrev2}
%%%%%%%%%%%%%%%%%% END OF BIBLIOGRAPHY USING BIBTEX%%%%%%%%%%%%%%%%%

\end{document}